\documentclass[preprint,prb,preprintnumbers,amsmath,amssymb,showpacs,floatfix]{revtex4}

\usepackage{graphicx,bm}

\makeatletter
\def\graphicscale{\twocolumn@sw{0.28}{0.35}}
\def\graphicthreescale{\twocolumn@sw{0.28}{0.35}}
\makeatother

\begin{document}

\title{Ferromagnetic-glassy transitions in three-dimensional Ising spin glasses}

\author{Giacomo Ceccarelli,$^{1}$ 
Andrea Pelissetto,$^2$ and Ettore Vicari$^1$ } 
\address{$^1$ Dip. Fisica dell'Universit\`a di Pisa and
INFN, Largo Pontecorvo 2, I-56127 Pisa, Italy} 
\address{$^2$
Dip. Fisica dell'Universit\`a di Roma ``La Sapienza" and INFN, P.le
Moro 2, I-00185 Roma, Italy} 

\date{July 13, 2011}

\begin{abstract}
  We investigate the ferromagnetic-glassy transitions which separate the
  low-temperature ferromagnetic and spin-glass phases in the
  temperature-disorder phase diagram of three-dimensional Ising spin-glass
  models. For this purpose, we consider the cubic-lattice $\pm J$
  (Edwards-Anderson) Ising model with bond distribution $P(J) = p \delta(J -
  1) + (1-p) \delta(J + 1)$, and present a numerical Monte Carlo study of the
  critical behavior along the line that marks the onset of ferromagnetism.
  
  The finite-size scaling analysis of the Monte Carlo data shows that
  the ferromagnetic-glassy transition line is slightly reentrant.  As
  a consequence, for an interval of the disorder parameter $p$, around
  $p\approx 0.77$, the system presents a low-temperature glassy phase,
  an intermediate ferromagnetic phase, and a high-temperature
  paramagnetic phase. Along the ferromagnetic-glassy transition line
  magnetic correlations show a universal critical behavior with
  critical exponents $\nu=0.96(2)$ and $\eta=-0.39(2)$.  The
  hyperscaling relation $\beta/\nu = (1 + \eta)/2$ is satisfied at the
  transitions, so that $\beta/\nu = 0.305(10)$. This magnetic critical
  behavior represents a new universality class for ferromagnetic
  transitions in Ising-like disordered systems. Overlap correlations
  are apparently not critical and show a smooth behavior across the
  transition.
\end{abstract}




\maketitle


\section{Introduction}
\label{intro}

Spin glass models are simplified, although still quite complex, models
retaining the main features of physical systems which show glassy behavior in
some region of their phase diagram.  They may be considered as theoretical
laboratories where the combined effects of disorder and frustration can be
investigated.  Their phase diagram and critical behavior can be used to
interpret the experimental results for complex materials.  Ising-like spin
glasses, such as the $\pm J$ Ising model,\cite{EA-75} model disordered
uniaxial magnetic materials characterized by random ferromagnetic and
antiferromagnetic short-ranged interactions, such as Fe$_{1-x}$Mn$_x$TiO$_3$
and Eu$_{1-x}$Ba$_x$MnO$_3$; see, e.g.,
Refs.~\onlinecite{IATKST-86,GSNLAI-91,NN-07}.  The random nature of the
short-ranged interactions is mimicked by nearest-neighbor random bonds.

Three-dimensional (3D) Ising spin glasses have been widely investigated. At
low temperatures they present ferromagnetic and glassy phases, depending on
the amount of frustration.  The critical behaviors along the finite-temperature
paramagnetic-ferromagnetic and paramagnetic-glassy (PG) transition lines have
been accurately
studied.\cite{HPPV-07-pf,HPV-08,KKY-06,Jorg-06,CHT-06,Betal-00,PC-99,KR-03} On
the other hand, the low-temperature behavior, in particular the nature of the
glassy phase and of the boundary between the ferromagnetic and glassy
phases, is still debated.

In this paper we focus on the low-temperature transition line which separates
the ferromagnetic phase, characterized by a nonzero magnetization, and the
spin-glass (glassy) phase in which the magnetization vanishes while the overlap
expectation value remains nonzero.  We consider the 3D $\pm J$ Ising model,
defined by the Hamiltonian\cite{EA-75}
\begin{equation}
H = - \sum_{\langle xy \rangle} J_{xy} \sigma_x \sigma_y,
\label{lH}
\end{equation}
where $\sigma_x=\pm 1$, the sum is over the nearest-neighbor sites of a cubic
lattice, and the exchange interactions $J_{xy}$ are uncorrelated quenched
random variables with probability distribution
\begin{equation}
P(J_{xy}) = p \delta(J_{xy} - 1) + (1-p) \delta(J_{xy} + 1).  
\label{pmjdi}
\end{equation}
The usual bimodal Ising spin glass model, for which $[J_{xy}]=0$
(brackets indicate the average over the disorder distribution),
corresponds to $p=1/2$.  For $p\neq 1/2$ we have $[J_{xy}]=2p-1\neq
0$, and ferromagnetic (or antiferromagnetic) configurations are
energetically favored.

\begin{figure}[tbp]
\includegraphics*[scale=\graphicscale]{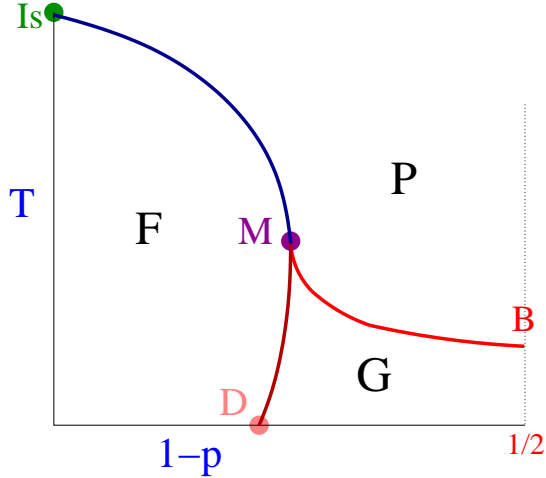}
\caption{ (Color online) Temperature-disorder phase
  diagram of the 3D $\pm J$ Ising model.  The phase diagram is
  symmetric under $p\to 1-p$ (but for small values of $p$ the 
  system is antiferromagnetic). }
\label{phadiad3}
\end{figure}

The phase diagram of the cubic-lattice $\pm J$ Ising model is sketched in
Fig.~\ref{phadiad3}.  We only consider $p\ge 1/2$ because of the symmetry
$p\to 1-p$.  While the high-temperature phase is always paramagnetic (P), at
low temperatures there is a ferromagnetic (F) phase for small frustration,
i.e., small values of $1-p$, and a glassy (G) phase with vanishing
magnetization for sufficiently large frustration.  In Fig.~\ref{phadiad3} we
do not report any low-temperature mixed phase with simultaneous glassy and
ferromagnetic behavior as found in mean-field models~\cite{Toulouse-80}, for
which, at present, there is no evidence.\cite{Hartmann-99,KM-02} The different
phases are separated by transition lines belonging to different universality
classes.  They meet at a magnetic-glassy multicritical point M located along
the so-called Nishimori line\cite{Nishimori-81,Nishimori-book}
$2/T=\ln[p/(1-p)]$, where the magnetic and the overlap two-point correlation
functions are equal.  Scaling arguments\cite{GHLB-85,HPPV-07-mcp} show that
the transition lines must be all parallel to the $T$ axis at the multicritical
point M.

The paramagnetic-ferromagnetic (PF) transition line starts at the
Ising transition of the pure system at $p=1$, at~\cite{H-10} $T_{\rm
Is}=4.5115232(16)$, with a correlation-length exponent $\nu_{\rm
Is}\approx 0.6301$ ($\nu=0.63012(16)$ from Ref.~\onlinecite{CPRV-02}
and $\nu=0.63002(10)$ from Ref.~\onlinecite{H-10}).  Along the PF line
the magnetic critical behavior is universal,\cite{HPPV-07-pf} and
belongs to the randomly-dilute Ising universality
class,\cite{ri-ft,ri-mc} characterized by the correlation-length
critical exponent $\nu_{\rm PF}=0.683(2)$.  It extends up to the
multicritical point M, located at\cite{HPPV-07-mcp} $p_{\rm
M}=0.76820(4)$, $T_{\rm M}=1.6692(3)$, whose multicritical behavior is
characterized by two even relevant renormalization-group (RG)
perturbations with RG dimensions $y_1 = 1.02(5)$ and $y_2 = 0.61(2)$.
The paramagnetic-glassy (PG) transition line runs from M to the
finite-temperature transition at $p=1/2$, at\cite{HPV-08} $T_{\rm
B}=1.11(1)$. The glassy critical behavior is universal along the PG
line;\cite{HPV-08} the overlap correlation-length exponent is quite
large,\cite{HPV-08,KKY-06,Jorg-06,CHT-06,Betal-00,PC-99} $\nu_{\rm
PG}=2.45(15)$.  Finally, (at least) another transition line is
expected to separate the ferromagnetic and glassy phases. This is the
ferromagnetic-glassy (FG) transition line that marks the onset of
ferromagnetism and which runs from M down to the point D at $T=0$.
The nature and the general features of this transition line in Ising
spin glasses are not known.  Beside a few numerical works at
$T=0$,\cite{Hartmann-99,KM-02} this issue has never been investigated
at finite temperature.

An interesting issue concerning the FG transition line is whether it
is reentrant, which would imply the existence of a range of values of
$p$ for which the glassy phase is separated from the paramagnetic
phase by an intermediate ferromagnetic phase.  As proved in
Refs.~\onlinecite{Nishimori-81,Nishimori-book}, ferromagnetism can
only exist in the region $p>p_{\rm M}$, which implies that $p_{\rm
D}\ge p_M$.  We also mention that, using {\em entropic} arguments
applied to frustration, the FG phase boundary was argued to run
parallel to the $T$ axis,\cite{Nishimori-86,Nishimori-book} i.e.,
$p_{\rm D}=p_{\rm M}$ for any $T<T_{\rm M}$, with the critical
behavior controlled by a $T=0$ {\em percolation} fixed
point.\cite{KR-03} The FG transition was numerically investigated at
$T=0$ in Ref.~\onlinecite{Hartmann-99}, obtaining the estimate $p_{\rm
D}=0.778(5)$ for the critical disorder, which is slightly larger than
$p_{\rm M} = 0.76820(4)$. Thus, it suggests a slightly reentrant FG
transition line, although its apparent precision is not sufficient to
exclude $p_{\rm D}=p_{\rm M}$.

In this paper we study the nature of the FG transition.  In
particular, we investigate whether the magnetic variables show a
continuous and universal critical behavior from M to D, and whether
hyperscaling is violated as it occurs in some systems whose critical
behavior is controlled by a zero-temperature fixed point, like the 3D
random-field Ising model.\cite{zeroTFP}

Note that we focus on the low-temperature ferromagnetic transition
line, which marks the onset of ferromagnetism moving from the glassy
phase with zero magnetization.  There is also the possibility that a
second low-temperature transition line exists for larger values of
$p$. In this case there would be a mixed low-temperature phase, in
which ferromagnetism and glassy order coexist. This occurs in
mean-field models~\cite{Toulouse-80} such as the infinite-range
Sherrington-Kirkpatrick model.\cite{SK-75} However, numerical $T=0$
ground-state calculations in the 3D $\pm J$ Ising model on a cubic
lattice\cite{Hartmann-99} and in related models\cite{KM-02} do not
seem to show evidence of a mixed phase and are consistent with a
unique transition.

In this paper we present a Monte Carlo (MC) study of the critical
behavior along the FG transition line. We perform simulations of
finite systems defined on cubic lattices of size $L\le 20$. A
finite-size scaling (FSS) analysis of numerical data at $T=0.5$ and
$T=1$ as a function of $p$ shows that magnetic correlations undergo a
continuous transition along the FG line. The critical behavior is
universal, i.e., independent of $T$ along the line. For the magnetic
critical exponents we obtain $\nu=0.96(2)$ and
$\eta=-0.39(2)$. Moreover, hyperscaling is verified.  The FG
transition line turns out to be slightly reentrant.  Indeed, we find
$p_c=0.7729(2)$ at $T=0.5$ and $p_c=0.7705(2)$ at $T=1$, which are
definitely larger than the disorder parameter $p_M=0.76820(4)$ at the
multicritical point.  Therefore, for a small interval of the disorder
parameter, around $p\approx 0.77$, the phase diagram presents three
different phases: a low-temperature glassy phase, an intermediate
ferromagnetic phase, and a high-temperature paramagnetic phase.

Note that the critical behavior of the magnetic correlations along the
FG transition line shows a new universality class of ferromagnetic
transitions in Ising-like disordered systems, which differs from the
randomly-dilute Ising universality class describing the critical
behavior along the PF transition line, and from the random-field
Ising universality class characterized by hyperscaling violation.

The general features of the phase diagram presented in
Fig.~\ref{phadiad3} should also characterize the temperature-disorder
phase diagram of other 3D Ising spin glass models with tunable
disorder parameters. For example, one may consider models with
Gaussian bond distributions, such as
\begin{equation}
P(J_{xy})\sim \exp\left[-{(J_{xy}-J_0)^2\over 2\sigma}\right],
\label{gauss}
\end{equation}
where the parameters $J_0$ and $\sigma$ control the amount of disorder
(the pure ferromagnetic model corresponds to $J_0>0$ and
$\sigma=0$). This distribution is also characterized by the presence
of a Nishimori line $T=\sigma/J_0$, where the magnetic and the overlap
two-point correlation functions are equal.  We also mention that an
analogous temperature-disorder phase diagram, with three transition
lines meeting at a multicritical point like Fig.~\ref{phadiad3}, is
also found in 3D XY gauge glass models.\cite{AV-11} A similar phase
diagram is also expected for other continuous spin glasses, like XY
and Heisenberg spin glasses with bond distributions (\ref{pmjdi}) or
(\ref{gauss}).

The paper is organized as follows. In Sec.~\ref{MCsim} we describe the
MC simulations, and provide the definitions of the quantities we
consider.  Sec.~\ref{numres} presents the FSS analysis
of the MC data, reporting the main results of the paper. Finally, in
Sec.~\ref{conclusions} we draw our conclusions.  In the appendix we
report some details of the FSS analyses.

\section{Monte Carlo simulations and observables}
\label{MCsim}

In order to study the FG transition line, which connects points M and
D in Fig.~\ref{phadiad3}, we perform MC simulations of the $\pm J$
Ising model on cubic lattices of size $L$ with periodic boundary
conditions.  We use the Metropolis algorithm, the random-exchange
method, and multispin coding. Implementation details can be found in
Ref.~\onlinecite{HPV-08}. In the random-exchange simulations we
consider $N_T$ systems at the same value of $p$ and at different
temperatures in the range $T_{\rm max} \ge T_i \ge T_{\rm min}$, with
$T_{\rm max}\gtrsim 2$ and $T_{\rm min}=0.5$.  The value $T_{\rm max}$
is chosen so that the thermalization at $T_{\rm max}$ is sufficiently
fast---typically we take $T_{\rm max}\gtrsim T_M\approx 1.67$---while
the intermediate values $T_i$ are chosen such that the acceptance
probability for the temperature exchange is at least $10\%$.  We
require one of the $T_i$ to be along the Nishimori
line.\cite{Nishimori-81} The results for this temperature value can be
compared with the known exact results and thus provide a check of the
MC code and the thermalization.  Finally, one of the temperatures
always corresponds to $T=1$.  The parameter $N_T$ increases with $L$
and varies from $N_T=5$ for $L=4$ to $N_T=19$ for $L=20$.
Thermalization is checked by verifying that disorder averages are
stable when increasing the number of MC steps for each disorder
realization.  We average over a large number $N_s$ of disorder
samples: $N_s \approx 2 \times 10^6$ samples for $L=4,6,8$,
$N_s\approx 3 \times 10^5$ for $L=10$, $N_s\approx 10^5$ for $L=12$,
$N_s\approx 5\times 10^4$ for $L=16$, and $N_s\approx 5\times 10^3$
for $L=20$.

The simulations are quite costly, because of the very slow dynamics
for low temperatures. This makes the computational effort increase
with a large power of the lattice size. In our range of values of $L$,
the number of iterations which must be discarded for thermalization
apparently increases as $L^{8}$ for our largest lattices (with an
increasing trend with increasing $L$).  Hence, taking into account the
volume factor, the CPU time for each disorder realization apparently
increases as $L^{11}$ (but we should warn that its large-$L$
asymptotic behavior may be even worse).  In total, simulations took
approximately 40 years of CPU time on a single core of a recent
standard commercial processor.

We consider the magnetization and the magnetic correlation function defined as
\begin{eqnarray}
&&m = {1\over V} {[\langle |\sum_x\sigma_x|\rangle]},\label{mGdef}\\
&&G(x) \equiv [ \langle \sigma_0 \sigma_x \rangle ],
\nonumber
\end{eqnarray}
where the angular and the square brackets indicate the thermal and the
quenched average over disorder, respectively. We define the magnetic
susceptibility and the second-moment correlation length, respectively as
\begin{eqnarray}
&&\chi\equiv \sum_{x} G(x),\label{xidefffxy}\\
&&\xi^2 \equiv  {1\over 4 \sin^2 (q_{\rm min}/2)} 
{\widetilde{G}(0) - \widetilde{G}(q)\over \widetilde{G}(q)},
\nonumber
\end{eqnarray}
where $q = (q_{\rm min},0,0)$, $q_{\rm min} \equiv 2 \pi/L$,
and $\widetilde{G}(q)$ is the Fourier transform of $G(x)$.  
Moreover, we consider the cumulants
\begin{eqnarray}
&&U_{4}  \equiv { [ \mu_4 ]\over [\mu_2]^{2}}, \label{Rdef}\\
&&U_{22} \equiv  {[ \mu_2^2 ]-[\mu_2]^2 \over [\mu_2]^2},\nonumber 
\end{eqnarray}
where
\begin{equation}
\mu_k\equiv\langle(\sum_x\sigma_x)^k\rangle.
\end{equation}
At the critical point $R_\xi\equiv \xi/L$, $U_4$, and $U_{22}$ (in the
following we call them phenomenological couplings and denote them by $R$) are
expected to approach universal values in the large-$L$ limit (within cubic
$L^3$ systems with periodic boundary conditions).  In the ferromagnetic phase
we have $U_4\to 1$, $U_{22}\to 0$, and $R_\xi\to\infty$, while in the glassy
phase we expect $R_\xi\to 0$.

We also define analogous quantities using the
overlap variables $q_x \equiv \sigma_x^{(1)} \sigma_x^{(2)}$, where
$\sigma_x^{(1)}$ and $\sigma_x^{(2)}$ are two 
independent replicas corresponding to the same couplings $J_{xy}$. 
In particular, we consider $\xi_o$ and $U_4^o$ 
defined by replacing the magnetic variables with the overlap variables
in Eqs.~(\ref{xidefffxy}) and (\ref{Rdef}).

\section{Finite-size scaling analysis}
\label{numres}

In this section we present a finite-size scaling (FSS) analysis of the MC data 
close to the FG transition line. We consider two values of the temperature,
$T = 0.5$ and $T = 1$, below the temperature
$T_M=1.6692(3)$ of the multicritical point M, and perform a FSS analysis as
a function of $p$.

\subsection{Phenomenological couplings and universality}
\label{fssanmo}

\begin{figure}[tbp]
\includegraphics*[scale=\graphicscale]{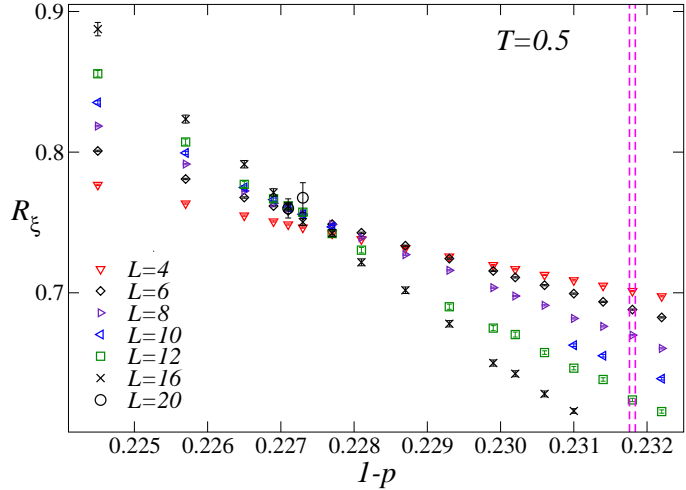}
\caption{(Color online) Estimates of $R_\xi$ at $T=0.5$.  The vertical
lines show the location of the multicritical point
M: $1-p_M=0.23180(4)$.}
\label{rxidata}
\end{figure}

To begin with, we analyze the data at $T=0.5$.  In Fig.~\ref{rxidata} we show
the MC estimates of $R_\xi$ as a function of $1-p$.  Analogous plots are
obtained for $U_4$ and $U_{22}$.  The data for different lattice sizes clearly
show crossing points, providing evidence for a continuous transition.  They
cluster at values of $p$ which are definitely larger than $p_M$, ruling out a
vertical transition line from M to the $T=0$ axis.

\begin{figure}[tbp]
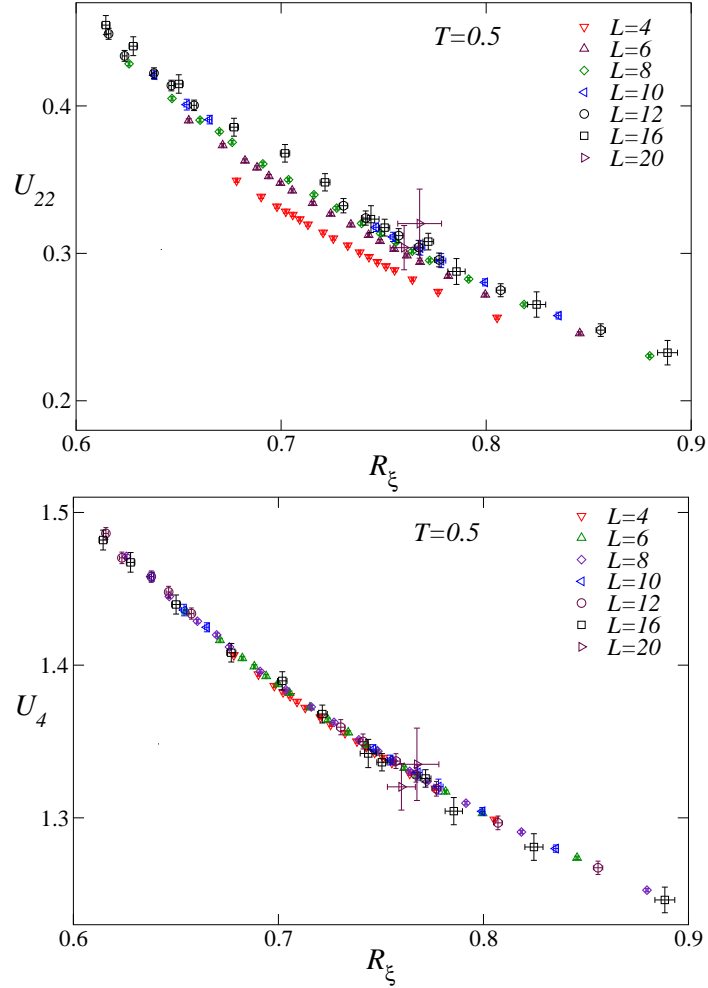

\includegraphics*[scale=\graphicscale]{rxiu22t0p5.eps}
\includegraphics*[scale=\graphicscale]{rxiu4t0p5.eps}
\caption{(Color online) $U_{4}$ (bottom) and $U_{22}$ (top)
vs $R_\xi$ at $T=0.5$.  }
\label{rxiut0p5}
\end{figure}

In the critical limit, the phenomenological couplings $R$ scale as 
\begin{equation}
    R = f_R[(p-p_c) L^{1/\nu}],
\label{R-FSS}
\end{equation}
where we have neglected analytic and nonanalytic scaling corrections.
Equivalently, one can test FSS by considering two different couplings $R_1$
and $R_2$. In the FSS limit $R_1 = F_{12}(R_2)$, where the function
$F_{12}(R_2)$ is universal, i.e., identical in any model that belongs to a
given universality class.  Clear evidence of FSS is provided in
Fig.~\ref{rxiut0p5}, where the phenomenological couplings $U_4$ and $U_{22}$
are reported versus $R_\xi\equiv \xi/L$.  
The data appear to rapidly approach a nontrivial
limit with increasing the lattice size. Scaling corrections are only visible
in the case of $U_{22}$, but they decrease with increasing $L$.

In order to determine the critical parameter $p_c$ and the 
exponent $\nu$, we fit $U_4$, $U_{22}$, and $R_\xi\equiv \xi/L$ to 
Eq.~(\ref{R-FSS}). Details are reported in App.~\ref{AppA}. We obtain
\begin{eqnarray}
&&p_c(T=0.5) = 0.7729(2), \qquad \nu=0.96(2),
\label{fgexp}\\
&&R_\xi^*=0.764(6), \quad U_4^*=1.331(5), \quad U_{22}^*=0.305(2), 
\label{Rexp}
\end{eqnarray}
where
$R^* = f_R(0)$ is the value of the phenomenological coupling $R$
at the critical point.
Scaling corrections turn out to be small. 

\begin{figure}[tbp]
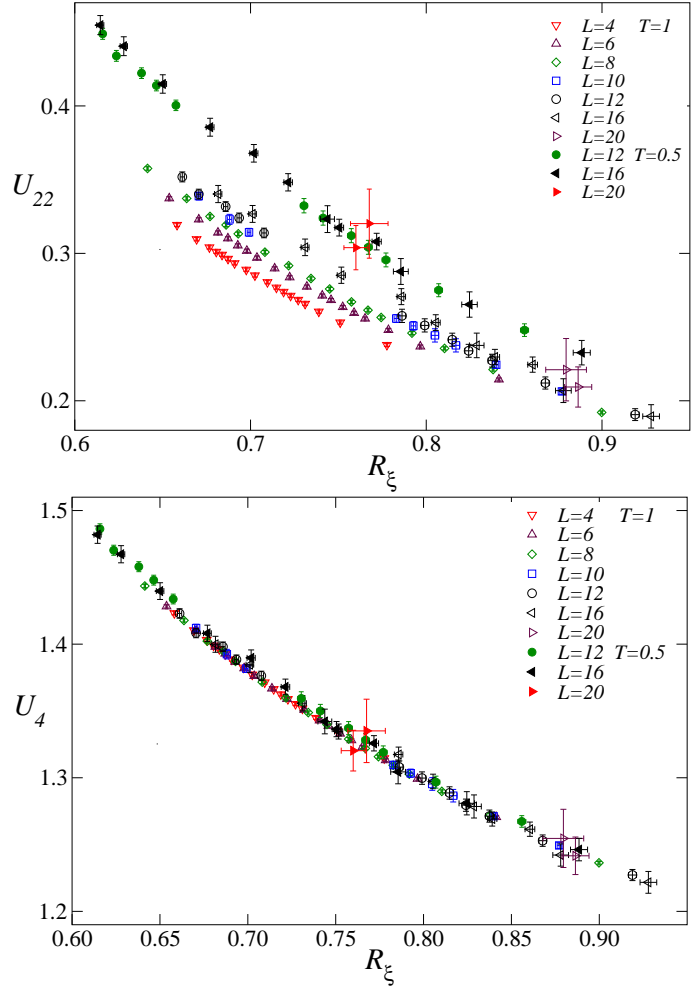

\includegraphics*[scale=\graphicscale]{rxiu22.eps}
\includegraphics*[scale=\graphicscale]{rxiu4.eps}
\caption{(Color online) $U_{4}$ (bottom) and $U_{22}$ (top)
vs $R_\xi$ at $T=1$ and at $T=0.5$ (only data with $L\ge 12$). }
\label{rxiu}
\end{figure}

An analogous FSS analysis can be performed at $T = 1$, with the 
purpose of checking universality, i.e., of determining whether all
transitions along the FG line belong to the same universality class.
For this purpose, we use the fact that, given any pair of RG
invariant quantities $R_1$ and $R_2$, the FSS function $R_1 = F_{12}(R_2)$
is universal. In
Fig.~\ref{rxiu} we plot $U_4$ and $U_{22}$ versus $R_\xi$ for both
$T=0.5$ and $T=1$.  The plot of $U_4$ provides good evidence of universality:
all data fall onto a single curve with remarkable precision. The results 
for $U_{22}$ show instead significant scatter, but they are also consistent with
universality if one takes into account scaling corrections: indeed, as $L$
increases the data for $T = 1$ approach the $T=0.5$ results.

For a more quantitative check, we must explicitly take into account 
scaling corrections at $T=1$, since they are significantly larger
than those observed at $T=0.5$. For instance, 
fits of the phenomenological couplings at $T=1$
to Eq.~(\ref{R-FSS}) show a somewhat large $\chi^2$/DOF (DOF is the 
number of degrees of freedom of the fit). Moreover, the estimates 
show systematic trends as the lattices with smaller values of $L$
are discarded in the fit, see App.~\ref{AppA} for details. 
To include scaling corrections, we fit the data to
\begin{equation}
    R = f_R[(p-p_c) L^{1/\nu}] + L^{-\omega} g_R[(p-p_c) L^{1/\nu}].
\end{equation}
The smallest $\chi^2$/DOF is obtained for 
$0.8\lesssim \omega \lesssim 0.9$. Correspondingly $\nu = 0.91(3)$, 
in substantial agreement with the estimate (\ref{fgexp}). 
Also the estimates of $R_\xi^*$, $U_4^*$,
and $U_{22}^*$, see App.~\ref{AppA},
are in agreement with the estimates (\ref{Rexp}) at $T=0.5$.
Therefore, all results 
strongly support the universality of the critical
behavior along the FG line. It is difficult to estimate reliably the 
exponent $\omega$ from the data. It we assume universality and 
fit the results at $T=1$ fixing $\nu = 0.96(2)$, we obtain 
$\omega = 0.95(10)$.
Note that the fits of the data at $T=0.5$
give much larger values for $\omega$, i.e., $\omega\gtrsim
2$, see App.~\ref{AppA}. This is probably due to the fact that 
corrections with $\omega\approx 1$ have very small amplitudes at $T=0.5$, 
so that we are simply measuring an effective exponent that mimicks the 
behavior of several correction terms.

The FSS fits also provide estimates of $p_c$ at $T=1$. We obtain 
\begin{equation}
p_c(T=1) = 0.7705(2).
\label{fgexpt1}
\end{equation}
Note that $p_c(T=1) > p_M\approx 0.7682$,
conferming the reentrant nature of the 
FG transition line.

\subsection{Magnetic susceptibility}

\begin{figure}[tbp]
\includegraphics*[scale=\graphicscale]{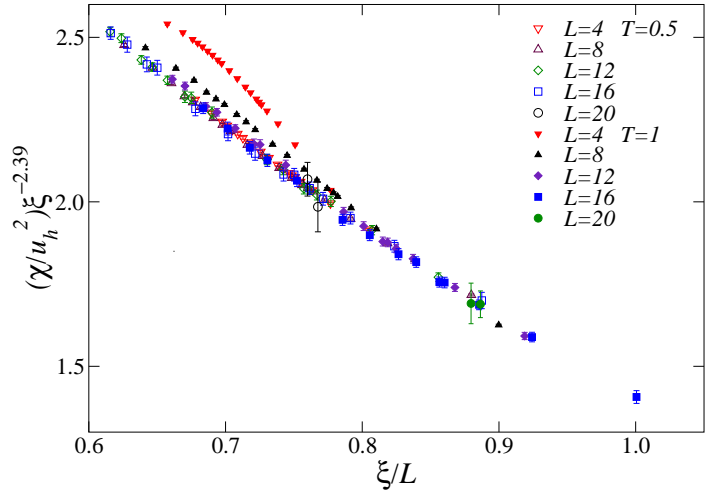}
\caption{(Color online) $\widetilde{\chi}\equiv
\chi u_h^{-2} \xi^{-2.39}$ versus 
$\xi/L$ for $T=1$ and $T=0.5$. }
\label{chisc}
\end{figure}

As discussed at length in Ref.~\onlinecite{HPV-08}, in the critical limit the 
magnetic susceptibility scales as 
\begin{equation}
\chi(p,L) =  u_h(p)^2 L^{2-\eta} f_\chi[(p-p_c) L^{1/\nu}],
\label{chi-sc1}
\end{equation}
where $u_h(p)$ is related to the magnetic scaling field and is 
an analytic function of $p$ (and also of the temperature). 
Fits of 
$\chi$ at $T=1$ and $T=0.5$ are good ($\chi^2$/DOF of order 1)
if we include all data such that $L\ge 6$, provided that 
$u_h(p)$ is taken into account (see App.~\ref{AppB} for details). 
We end up with the final estimate
\begin{equation}
\eta=-0.39(2).
\label{etaest}
\end{equation}
Since $\xi/L$ is a function of $(p-p_c) L^{1/\nu}$ in the FSS limit, 
see Eq.~(\ref{R-FSS}), we can rewrite Eq.~(\ref{chi-sc1}) as 
\begin{equation}
\chi(p,L) =  u_h(p)^2 \xi^{2-\eta} F_\chi(\xi/L).
\label{chi-sc2}
\end{equation}
The function $F_\chi(x)$ is universal apart from a multiplicative constant,
which takes into account the freedom in the normalization of the 
function $u_h(p)$. In Fig.~\ref{chisc} we show the quantity
$\widetilde{\chi} = \chi u_h^{-2} \xi^{-2.39} $ for $T=1$ and $T=0.5$. 
For each temperature the function $u_h(p)$ is determined by fitting
the susceptibility data to Eq.~(\ref{chi-sc2}), fixing $\eta = -0.39$. 
Moreover, the scaling fields are 
normalized so that $\widetilde{\chi}(T=1,L=16) \approx 
\widetilde{\chi}(T=0.5,L=16)$ for $\xi/L \approx 0.8$. If we discard the data 
with $L=4$ and 8 at $T=0.5$, all points fall on top of each other, 
confirming universality.

\subsection{Evidence of hyperscaling}
\label{hyper}

\begin{figure}[tbp]
\includegraphics*[scale=\graphicscale]{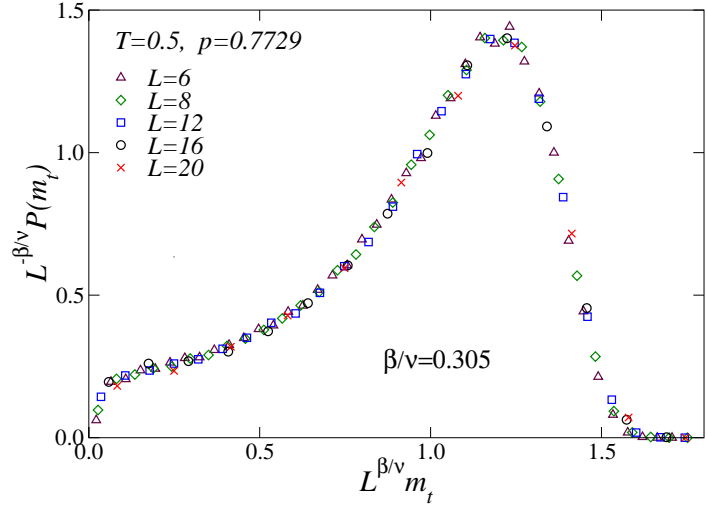}
\caption{(Color online) Scaling behavior of the 
distribution of the thermal averages of the magnetization,
at $T=0.5$ and $p=p_c=0.7729$. We set
$\beta/\nu=0.305$.  }
\label{hyst}
\end{figure}

Since the FG transition line extends up to $T=0$, hence the 
critical behavior may be controlled by a zero-temperature fixed point, 
hyperscaling might be violated, as it happens 
in the 3D random-field Ising model.\cite{zeroTFP}
In order to check whether hyperscaling holds along the FG line, we consider
the magnetization, which is expected to behave as $m\sim L^{-\beta/\nu}$
at the critical point,
and the magnetic susceptibility, which scales as $\chi\sim L^{2-\eta}$.
If hyperscaling holds, $\beta$ and $\eta$ are related by 
\begin{equation}
{\beta\over \nu} = {d-2+\eta\over 2},
\label{hypersc}
\end{equation}
(in the present case $d=3$), which guarantees that $\chi/m^2$ scales as $L^d$. 
In order to verify whether Eq.~(\ref{hypersc}) holds, 
we consider $H\equiv\chi/(m^2 L^3)$ and assume 
that it behaves as 
\begin{equation}
H \equiv {\chi\over m^2 L^3} \sim L^{\zeta} f_H[(p-p_c) L^{1/\nu}].
\label{chiom2}
\end{equation}
If hyperscaling holds, $\zeta$ vanishes.  A FSS analysis of the data
at $T=0.5$ and $T=1$ gives the rather stringent bound (details in
App.~\ref{AppC}) 
\begin{equation}
|\zeta| < 0.01, 
\label{zetaest}
\end{equation}
which allows us to conclude, quite confidently,
that hyperscaling holds. If this the case, using estimates
(\ref{etaest}) and (\ref{fgexp}) of $\eta$ and $\nu$, we obtain
\begin{equation}
\beta/\nu=(1+\eta)/2 = 0.305(10), \qquad 
\beta = 0.29(1).
\label{betaest}
\end{equation}
As a further check, we consider the sample distribution $P(m_t)$ of
the thermal averages of the magnetization
\begin{equation}
m_t \equiv {1\over V} {\langle |\sum_x\sigma_x|\rangle}
\label{mtdef}
\end{equation}
at the critical point $p=p_c=0.7729$, $T=0.5$, 
which is expected to behave asymptotically as
\begin{equation}
P(m_t) \approx L^{\beta/\nu} {\cal P}(L^{\beta/\nu}m_t).
\label{pmt}
\end{equation}
In Fig.~\ref{hyst} we plot ${\cal P}(L^{\beta/\nu}m_t)$ using 
$\beta/\nu=0.305$. The data clearly show the expected scaling behavior.
In conclusion, the numerical results do not show evidence of 
hyperscaling violations in the critical behavior of magnetic correlations.

\begin{figure}[tbp]
\includegraphics*[scale=\graphicscale]{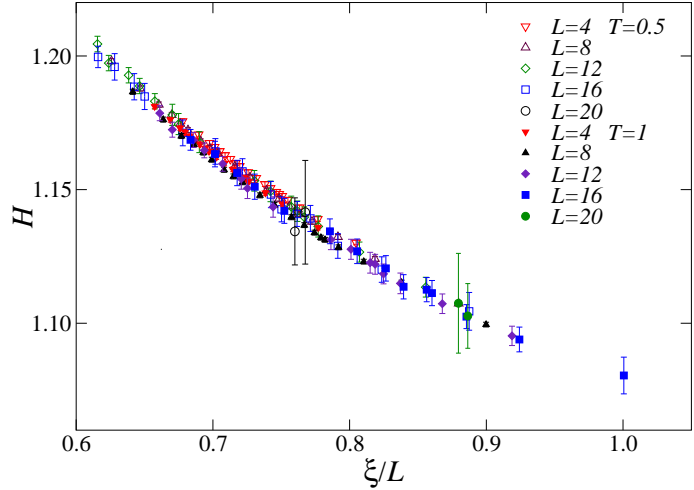}
\caption{(Color online) 
$H\equiv\chi/(m^2 L^3)$ versus $\xi/L$ 
for $T=1$ and $T=0.5$. }
\label{hyperscfig}
\end{figure}

Our data for $H(p,L)$ can also be used to provide further evidence of
universality.  Indeed, if we use the fact that $\xi/L$ is a function
of $(p-p_c) L^{1/\nu}$, see Eq.~(\ref{R-FSS}), we can rewrite
Eq.~(\ref{chiom2}) for $\zeta = 0$ as
\begin{equation}
   H(p,L) = F_H(\xi/L) + O(L^{-\omega}),
\end{equation}
where $F_H(x)$ should be the same at $T=0.5$ and at $T = 1$ if 
all transitions along the FG transition line belong to 
the same universality class. The plot of the data, see Fig.~\ref{hyperscfig}, 
clearly confirms universality: all points fall onto a single curve.

\subsection{Overlap correlations}
\label{ocorr}

\begin{figure}[tb]
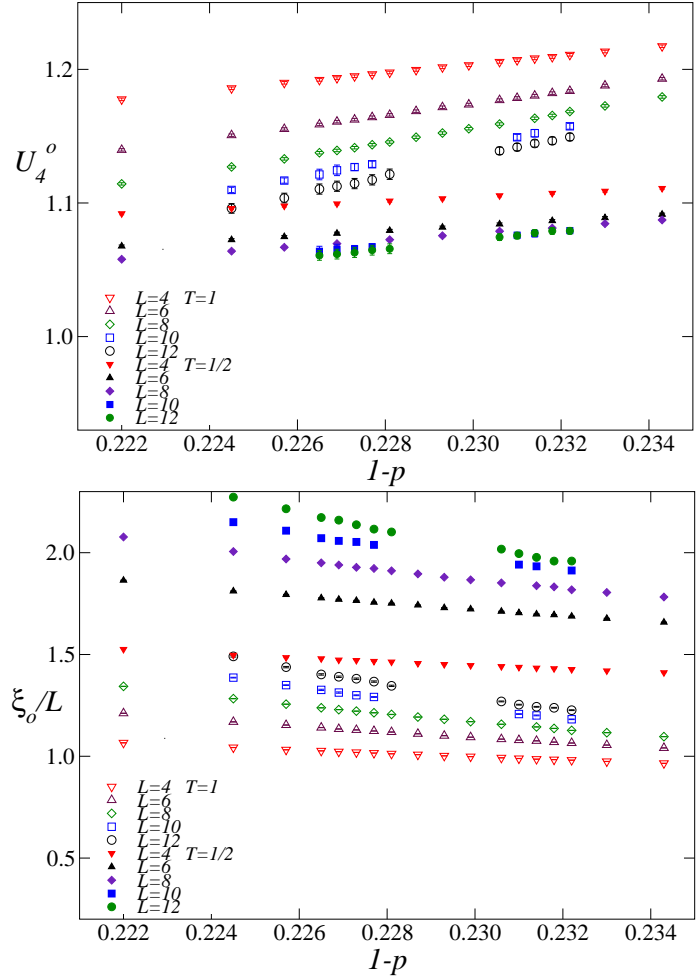

\includegraphics*[scale=\graphicscale]{u4o.eps}
\includegraphics*[scale=\graphicscale]{rxio.eps}
\caption{(Color online) Estimates of $\xi_o/L$ 
(bottom) and $U_4^o$ (top), defined
  in terms of the overlap variables, at $T=0.5$ and $T=1$.  }
\label{rxio}
\end{figure}

In our numerical study we also consider quantities involving the
overlap variables, such as $\xi_o/L$ and $U_4^o$, defined at the end
of Sec.~\ref{MCsim}.  In Fig.~\ref{rxio} we show MC data up to $L=12$
(since their computation turned out to be significantly more
demanding, we restricted the measurements for the lattices $L=16,\,20$
to the magnetic correlations).  Unlike the magnetic quantities, the
overlap data  do not show crossings in the interval of $p$ we have
investigated.  Apparently $U_4^o$ decreases continuously, while
$\xi_o/L$ increases as $L\to \infty$.  This may reflect the fact that
the FG transition line separates two {\em ordered} phases with respect
to the overlap variables. Note that the differences between data at
the same $p$ and $T$ and at different values of $L$ decrease as $1-p$
increases.  Hence, if there is a line in the $(p,T)$ plane where the
overlap variables show crossings, it must be such that $1-p > 0.234$,
i.e., it must lie in the region $p < p_M$, where no ferromagnetism is
possible.

\section{Conclusions}
\label{conclusions}

We investigate the critical behavior along the ferromagnetic-glassy
transition line of the $T$-$p$ phase diagram of the cubic-lattice $\pm
J$ (Edwards-Anderson) Ising model, cf.  Eq.~(\ref{lH}), which marks
the low-temperature boundary between the ferromagnetic phase and the
glassy phase where the magnetization vanishes,
i.e., the transition line that runs from M
down to the point D at $T=0$ in Fig.~\ref{phadiad3}.

We present a numerical study based on MC simulations of systems of
size up to $L=20$, obtaining MC estimates of several quantities at
$T=0.5$ and $T=1$ (which are well below the temperature
$T_M=1.6692(3)$ of the multicritical point M) as a function of the
disorder parameter $p$.  The results of the FSS analyses are consistent
with the two continuous magnetic transitions 
belonging to the same universality class. The corresponding 
critical exponents are $\nu=0.96(2)$ and $\eta=-0.39(2)$.  Since the
critical line extends up to $T=0$, the critical behavior may be
controlled by a zero-temperature fixed point. Correspondingly, it is
possible to have hyperscaling violations, as it occurs in the 3D
random-field Ising model. Our MC results show that the hyperscaling
relation $\beta/\nu = (1 + \eta)/2$ is satisfied, so that $\beta/\nu =
0.305(10)$ and $\beta=0.29(1)$.  
The FSS results provide a robust evidence of a universal
magnetic critical behavior along the FG transition line.  A reasonable
hypothesis is that also the zero-temperature transition belongs to the
same universality class. This is supported by the available numerical
data at $T=0$.  The numerical study of Ref.~\onlinecite{Hartmann-99}
for the $\pm J$ Ising model at $T=0$, using lattice sizes up to
$L=14$, provided evidence of a magnetic transition at $p_{\rm
D}=0.778(5)$, with critical exponents $\nu=1.3(3)$ and $\beta=0.2(1)$.
Numerical analyses\cite{KM-02} for other Ising spin-glass models at
$T=0$ give consistent values of the critical exponents, $\nu=0.9(2)$
and $\beta=0.3(1)$ using data up to $L=12$.  These estimates are
substantially consistent with our results along the FG transition
line, supporting a universal critical behavior along the FG transition
from the multicritical point M down to the $T=0$ axis.

We also investigate the behavior of overlap correlations. They 
do not appear to be critical and show an apparently smooth
behavior across the FG transition.  Our numerical results do not show
evidence of other transitions close to the transition line where ferromagnetism
disappears. Thus, they do not hint at the existence of a mixed
ferromagnetic-glassy phase, as found in mean-field models,\cite{Toulouse-80}
in agreement with earlier $T=0$ numerical studies.\cite{Hartmann-99,KM-02}

The FG transition line is slightly reentrant.  Indeed, we
find that $p_c=0.7729(2)$ at $T=0.5$ and
$p_c=0.7705(2)$ at $T=1$, which are definitely larger than
$p_M=0.76820(4)$, although they are quite close.  This implies that
there exists a small interval of the disorder parameter, around
$p\approx 0.77$, showing three different phases when varying $T$: with
increasing the temperature, the system goes from the low-temperature glassy 
phase with zero magnetization,
to an intermediate ferromagnetic phase, and finally to the 
high-temperature
paramagnetic phase.  Correspondingly, it first undergoes 
a glassy-ferromagnetic
transition with $\nu=0.96(2)$ and then a ferromagnetic-paramagnetic
transition with $\nu=0.683(2)$.  We mention that a slightly reentrant
low-temperature transition line, where ferromagnetism disappears, also
occurs in the phase diagram of the 2D $\pm J$ Ising
model.\cite{PPV-09,PHP-06}

The main features of the FG transition line are not expected to depend
on the particular discrete bond distribution of the $\pm J$ Ising
model, cf. Eq.~(\ref{pmjdi}). They should also apply to more general
distributions with tunable disorder parameters, such as the Gaussian
distribution reported in Eq.~(\ref{gauss}), and also to experimental
spin glass systems with tunable disorder.

\begin{figure}[tbp]
\includegraphics*[scale=\graphicscale]{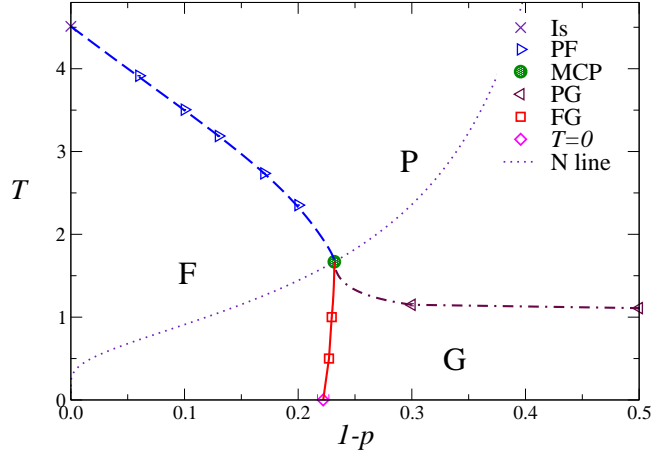}
\caption{(Color online) 
Numerical results for the phase boundaries of the cubic-lattice
$\pm J$ Ising model (\ref{lH}) in the $T$-$p$ plane.
The dashed lines are interpolations of the data~\cite{footnotemcp}.
}
\label{tp}
\end{figure}

We conclude showing Fig.~\ref{tp} which reports all available
numerical results for the phase boundaries of the cubic-lattice $\pm
J$ Ising model (\ref{lH}) in the $T$-$p$ plane, taken from
Ref.~\onlinecite{HPPV-07-pf} for the PF transition line, from
Ref.~\onlinecite{HPPV-07-mcp} for the multicritical point along the
Nishimori (N) line $T=2/{\rm ln}[p/(1-p)]$, from
Ref.~\onlinecite{HPV-08} for the data along the PG line, from this
paper along the FG line, and from Ref.~\onlinecite{Hartmann-99} for
the $T=0$ transition point.  The dashed lines are interpolations of
the data along the transition lines which satisfy the expected scaling
behavior at the multicritical point where they meet, controlled by the
crossover exponent $\phi=1.67(10)$, see
Refs.~\onlinecite{HPPV-07-mcp,PPV-09} for details.~\cite{footnotemcp}

\acknowledgements The MC simulations were performed at the INFN Pisa
GRID DATA center, using also the cluster CSN4.  

\appendix

\section{Analysis details}
\label{fitdetails}

\subsection{Phenomenological couplings}  \label{AppA}

\begin{table}
\caption{Results of combined fits 
of $U_4$, $U_{22}$, and $R_\xi$ to Eq.~(\ref{scaling-R}) without scaling 
corrections. $\chi^2$ is the sum of the residuals in the fit and DOF is 
the number of degrees of freedom. Each column corresponds to results in
which only data satisfying $L\ge L_{\rm min}$ are included.
$R^* \equiv f_R(0)$ is the value of the phenomenological coupling at the 
critical point.
}
\label{table-R}
\begin{tabular}{lcccc}
\hline\hline
\multicolumn{5}{c}{$T = 0.5$}  \\
\multicolumn{1}{l}{$L_{\rm min}$} &
\multicolumn{1}{c}{4} &
\multicolumn{1}{c}{6} &
\multicolumn{1}{c}{8} &
\multicolumn{1}{c}{10} \\
\hline
$\chi^2$/DOF  & 5594/289  &  567/229   &   203/169  &   79/109   \\
$\nu$         & 0.971(4)  &  0.964(5)  &   0.954(8) &   1.00(2)  \\
$p_c$         & 0.77230(1)&  0.77275(2)&  0.77284(3)&  0.77281(5)\\
$R_\xi^*$     & 0.7453(2) &  0.7564(3) &   0.7592(6)&   0.759(2) \\
$U_4^*$       & 1.3450(2) &  1.3364(4) &   1.3343(6)&   1.334(2) \\
$U_{22}^*$    & 0.3046(2) &  0.3045(3) &   0.3057(6)&   0.310(2) \\
\hline
\multicolumn{5}{c}{$T = 1$}  \\
\multicolumn{1}{l}{$L_{\rm min}$} &
\multicolumn{1}{c}{4} &
\multicolumn{1}{c}{6} &
\multicolumn{1}{c}{8} &
\multicolumn{1}{c}{10} \\
\hline
$\chi^2$/DOF  & 10593/289 &  1842/229  &  365/169   &  98/109     \\
$\nu$         & 1.054(5)  &  0.995(5)  &  0.963(8)  &  0.982(20)  \\
$p_c$         & 0.76819(2)&  0.76920(2)&  0.76975(3)&  0.76994(5) \\
$R_\xi^*$     & 0.6826(2) &  0.7019(3) &  0.7147(6) &  0.7220(17) \\
$U_4^*$       & 1.3973(3) &  1.3779(4) &  1.3662(6) &  1.3613(19) \\
$U_{22}^*$    & 0.3094(3) &  0.3020(4) &  0.2977(5) &  0.3013(17) \\
\hline\hline
\end{tabular}
\end{table}

In order to determine the exponent $\nu$ and the critical parameter $p_c$, 
we analyze the phenomenological couplings $U_{4}$, $U_{22}$, and 
$R_\xi\equiv \xi/L$. In the critical limit each quantity $R$ behaves as 
\begin{equation}
R(p,L) \approx f_R[u_p(p) L^{1/\nu}] + u_\omega(p) L^{-\omega} 
         g_R[u_p(p) L^{1/\nu}],
\label{scaling-R}
\end{equation}
where the nonlinear scaling fields $u_p(p)$ and $u_\omega(p)$ 
are analytic functions of $p$. We have $u_p(p_c) = 0$ while,
in general, we expect $u_\omega(p_c) \not=0$. For both 
temperatures our data belong to a small interval of values of $p$,
so that we expect the approximations $u_p(p) \approx p - p_c$ and 
$u_\omega(p) \approx u_\omega(p_c) = a_\omega$ to work well. To check it,
we also performed fits assuming 
$u_p(p) = p - p_c + k (p-p_c)^2$. We did not 
find any significant difference.

We first analyze the results at $T = 0.5$. We perform combined fits of the 
three quantities to Eq.~(\ref{scaling-R}) without scaling corrections 
(we set $a_\omega = 0$). If the  
scaling functions $f_R$ are approximated by fourth-order polynomials, we 
obtain the results reported in Table ~\ref{table-R}. We report estimates for 
different $L_{\rm min}$: in each fit we only include data satisfying 
$L\ge L_{\rm min}$. Corrections are quite small and indeed the results 
corresponding to $L_{\rm min} = 8$ and $L_{\rm min} = 10$ mostly agree 
within errors. We also perform fits that take into account scaling corrections.
We fix $\omega$, approximate $g_R(x)$ by a second-order polynomial,
and repeat the fit for several values of $\omega$ between 1 and 5. 
If we perform a combined fit of $U_4$ and $R_\xi$ (we include all
results with $L\ge 4$), the smallest $\chi^2$/DOF
(DOF is the number of degrees of freedom of the fit) is obtained 
for $3\lesssim \omega \lesssim 4$ and one would estimate 
$\nu = 0.96(1)$ and $p_c = 0.7729(1)$. If instead we use $U_4$, $R_\xi$,
and also $U_{22}$ we obtain $\omega \approx 2$,
$\nu = 0.95(1)$, and $p_c = 0.7731(1)$. These results indicate that 
scaling corrections are quite small, and quite probably cannot be parametrized
be a single correction term. Our best estimates of $\omega$ are simply 
effective exponents that parametrize the contributions of several different
correction terms, which are all relevant for our small lattice sizes.

If we compare all results, we end up with the estimates 
$p_c = 0.7729(2)$ and $\nu=0.96(2)$, reported in Eq.~(\ref{fgexp}).
For the phenomenological couplings at criticality, $R^* \equiv f_R(0)$, we 
obtain the estimates reported in Eq.~(\ref{Rexp}), i.e.,
$R_\xi^*= 0.764(6)$, $U_4^*=1.331(5)$ and $U_{22}^*=0.305(2)$.
The final estimates and their errors take into account the results of the fits 
with and without scaling corrections.

The same analyses can be performed at $T = 1$. Combined fits 
to Eq.~(\ref{scaling-R}) without scaling corrections give the 
results reported in Table \ref{table-R}. 
It is quite clear that scaling corrections at $T = 1$ are larger 
then those at $T = 0.5$. The goodness of the fit is worse and the 
fit results show systematic trends. It is however reassuring that 
they apparently converge towards the estimates (\ref{fgexp}) and 
(\ref{Rexp}), in agreement with universality. 

It is interesting to check whether scaling corrections can explain the
differences which occur among the results for $T=1$ reported in
Table~\ref{table-R} and the results obtained at $T = 0.5$. Since the results
for $U_{22}^*$ at $T = 1$ are nonmonotonic as a function of $L_{\rm min}$, at
least two correction terms must be included to explain the observed trend of
the data. Therefore, the fit of the $U_{22}$ data with a single scaling
correction makes no sense. In any case, the estimate obtained for $L_{\rm min}
= 10$ differs from the one reported in Eq.~(\ref{Rexp}) by one combined
error bar, and therefore is in agreement with universality.  We then perform
combined fits of $U_4$ and $\xi/L$ to Eq.~(\ref{scaling-R}), approximating
$g_R(x)$ by a second-order polynomial and fixing $\omega$ to several values
between 0.5 and 1.5.  The smallest $\chi^2$/DOF is obtained for $0.8\lesssim
\omega \lesssim 0.9$.  Correspondingly, we obtain $p_c = 0.7705(1)$, $R_\xi^*
= 0.765(10)$, and $U_4^* = 1.32(1)$. The estimates of the phenomenological
couplings at criticality are now in very good agreement with the estimates at
$T = 0.5$.  As for $\nu$ we obtain $\nu = 0.91(3)$, which is sligthly smaller
than, but still consistent with the estimate at $T = 0.5$.  If we fix $\nu =
0.96(2)$ as obtained at $T=0.5$, we find $\omega = 0.95(10)$, $p_c =
0.7704(1)$, $R_\xi^* = 0.757(7)$, $U_4^* = 1.326(6)$.

These fits provide an estimate of $p_c$ at $T=1$. We quote the estimate
$p_c = 0.7705(2)$ already reported in Eq.~(\ref{fgexpt1}),
which satisfies the inequality $p_c \gtrsim 0.7700$, which one would obtain
from the results reported in Table~\ref{table-R}. It is unclear how reliable
our estimates of $\omega$ are. In any case, they suggest a value close to 1.

\subsection{Magnetic susceptibility} \label{AppB}

\begin{table}
\caption{Estimates of the exponent $\eta$ obtained by fits to
Eq.~(\ref{fit-chi}), where $\hat{f}_\chi$ is approximated by a fourth-order
polynomial and $\hat{u}(p)$ by a second-order polynomial. 
In each fit we only include
the data which satisfy $L\ge L_{\rm min}$. 
We fix $\nu = 0.96(2)$ and the value of $p_c$:
$p_c = 0.7729(2)$ at $T = 0.5$ and $p_c = 0.7705(2)$ at $T = 1$.  
}
\label{table-eta}
\begin{tabular}{lcccc}
\hline\hline
& \multicolumn{2}{c}{$T = 0.5$}  & \multicolumn{2}{c}{$T = 1$} \\
\multicolumn{1}{l}{$L_{\rm min}$} &
\multicolumn{1}{c}{$\chi^2$/DOF} &
\multicolumn{1}{c}{$\eta$} &
\multicolumn{1}{c}{$\chi^2$/DOF} &
\multicolumn{1}{c}{$\eta$} \\
\hline
4  &  516/94 &  $-0.414(6)$ &    62/94 & $-0.393(6)$   \\
6  &   39/74 &  $-0.400(8)$ &    18/74 & $-0.389(9)$   \\
8  &   22/54 &  $-0.397(12)$&    16/54 & $-0.389(12)$  \\
10 &   11/34 &  $-0.398(16)$&     6/34 & $-0.390(16)$  \\
\hline\hline
\end{tabular}
\end{table}

We analyze the magnetic susceptibility which should scale as 
\begin{equation}
  \chi(p,L) = u_h(p)^2 L^{2-\eta} f_\chi[u_p(p) L^{1/\nu}],
\end{equation}
where $u_h(p)$ is related to the magnetic scaling field and is an 
analytic function of $p$; scaling corrections have been
neglected. In order to determine $\eta$, we perform fits to 
\begin{equation}
\ln \chi = (2-\eta) \ln L + \hat{f}_\chi[(p-p_c) L^{1/\nu}] + 
            \hat{u}(p),
\label{fit-chi}
\end{equation}
where $\hat{f}_\chi$ is approximated by a fourth-order polynomial 
and $u_h(p)$ is normalized so that $\hat{u}(p=p_c) = 0$.  In this
expression we have replaced $u_p(p)$ with $p-p_c$. Inclusion of the
second-order term does not change the quality of the fit and the results.
Instead, even if the interval in $p$ is small, the function $u_h(p)$ cannot be
approximated by a constant, hence $\hat{u}(p)$ cannot be set to zero.  Indeed,
the fits in which $\hat{u}(p)$ is approximated by a second-order polynomial 
have a $\chi^2/\hbox{\rm DOF}$ which is significantly smaller than those in
which we set $\hat{u}(p) = 0$. For instance, for $T=0.5$ and $L_{\rm min} = 6$
(we fix $p_c$ and $\nu$, see caption of Table~\ref{table-eta}), we have
$\chi^2/\hbox{\rm DOF} = 265/76$ and 39/74 for the fit with $\hat{u}(p) = 0$
and the fit with a second-order polynomial, respectively.  The results of the
fits in which we fix $\nu$ and $p_c$ are reported in Table~\ref{table-eta}.
The results are very stable with $L_{\rm min}$ and are completely consistent
with universality. Note that, at variance with what is observed for the
phenomenological couplings, corrections for $T = 1$ are apparently smaller
than for $T = 0.5$. This may indicate the presence of several corrections
which cancel out for our values of $L$. A conservative final estimate is
$\eta = -0.39(2)$, already reported in Eq.~(\ref{etaest}).

\subsection{Hyperscaling} \label{AppC}

\begin{table}
\caption{Estimates of the exponent $\zeta$. In each fit we only include
the data which satisfy $L\ge L_{\rm min}$.
We fix $\nu = 0.96(2)$ and the value of $p_c$:
$p_c = 0.7729(2)$ at $T = 0.5$ and $p_c = 0.7705(2)$ at $T = 1$.  
}
\label{table-zeta}
\begin{tabular}{lcccc}
\hline\hline
& \multicolumn{2}{c}{$T = 0.5$}  & \multicolumn{2}{c}{$T = 1$} \\
\multicolumn{1}{l}{$L_{\rm min}$} &
\multicolumn{1}{c}{$\chi^2$/DOF} &
\multicolumn{1}{c}{$\zeta$} &
\multicolumn{1}{c}{$\chi^2$/DOF} &
\multicolumn{1}{c}{$\zeta$} \\
\hline
4  &  88/98  &   $-0.007(2)$ &   161/98   &  $-0.015(1)$ \\
6  &  9/78   &   $-0.003(2)$ &    15/78   &  $-0.009(2)$ \\
8  &  8/58   &   $-0.002(3)$ &     3/58   &  $-0.006(3)$ \\
10 &  4/38   &   $-0.005(5)$ &     2/58   &  $-0.005(5)$ \\
\hline\hline
\end{tabular}
\end{table}

In order to study hyperscaling we consider the ratio 
\begin{equation}
  H \equiv {\chi\over m^2 L^3}.
\end{equation}
If hyperscaling holds, it should behave as 
\begin{equation}
  H(p,L) = f_h[u_p(p) L^{1/\nu}] \approx f_H[(p - p_c) L^{1/\nu}],
\label{scaling-H}
\end{equation}
where we have neglected scaling corrections.
In order to allow for a possible hyperscaling violation we introduce
a new exponent $\zeta$ and assume that 
\begin{equation}
  H(p,L) = L^{\zeta} f_H[(p - p_c) L^{1/\nu}].
\end{equation}
To determine $\zeta$ we perform fits to 
\begin{equation}
  \ln H(p,L) = \zeta \ln L + \hat{f}_H[(p - p_c) L^{1/\nu}],
\end{equation}
where $\hat{f}_H(x)$ is approximated by a second-order polynomial. 
Fit results are reported in Table~\ref{table-zeta}. Here we 
fix $\nu$ and $p_c$ to the values determined above. The quality of the 
fits is very good and scaling corrections are apparently small for both
values of the temperature. The exponent $\zeta$ is clearly compatible with 
zero, proving that hyperscaling is satisfied. More precisely, 
we obtain the bound $ |\zeta| < 0.01$, already reported in
Eq.~(\ref{zetaest}).


\begin{thebibliography}{99}
\bibitem{footnotemcp} An interpolation of the PF data with the correct
scaling behavior at the multicritical point is provided by $p = 0.7682
+ (0.59909 - \beta)^\phi (2.70487 - 9.03122\beta + 9.70235 \beta^2)$,
with $\beta \equiv 1/T$ and $\phi = 1.67$.  In Fig.~\ref{tp} the FG
line with $T \ge 1$ is approximated by the line of
equation~\cite{PPV-09} $p-p_M = a u_2(p,T)^\phi$, where $a \approx
0.03$ (which is fixed using our numerical estimate of $p_c$ at $T=1$),
$u_2(p,T) = \tanh \beta - 2p+1$, and $p_M$ is the position of the
multicritical point.  For $T \le 1$ we report the straight lines
connecting the data at $T=1,\,0.5$ and $T=0.5,\,0$.  Analogously, we
proceed for the PG line, reporting the curve $p-p_M = c u_2(p,T)^\phi$
for $p \ge 0.7$ and a straight line for $0.5\le p\le 0.7$.
\end{thebibliography}
\end{document}